\documentclass[preprint,pteplogo]{ptephy_v2}

\preprintnumber{}

\usepackage{amsmath,amssymb,graphicx,booktabs,multirow}
\usepackage{xspace}
\usepackage{textcomp}
\usepackage{textgreek}
\usepackage{siunitx}
\usepackage{hyperref}
\hypersetup{hidelinks}
\usepackage{subcaption}

\usepackage{lineno}
\DeclareSIUnit{\mm}{mm}
\DeclareSIUnit{\cm}{cm}
\DeclareSIUnit{\mT}{mT}
\DeclareSIUnit{\kW}{kW}
\DeclareSIUnit{\GeV}{GeV}
\DeclareSIUnit{\MeVc}{MeV/$c$}
\DeclareSIUnit{\POT}{POT}
\DeclareSIUnit{\TPOT}{10^{12}\,POTs}

\newcommand{\fig}[1]{\figurename~\ref{fig:#1}}
\newcommand{\tab}[1]{\tablename~\ref{tab:#1}}
\newcommand{\eq}[1]{Eq.~\eqref{eq:#1}}
\newcommand{\sect}[1]{Section~\ref{sec:#1}}

\newcommand{\PhaseAlpha}{Phase-\textalpha\xspace}
\newcommand{\T}[1]{\ensuremath{\operatorname{T_{#1}}}}
\newcommand{\eff}[1]{\ensuremath{\epsilon_{\operatorname{#1}}}}
\newcommand{\pur}[1]{\ensuremath{\eta_{\operatorname{#1}}}}
\newcommand{\accep}[1]{\ensuremath{A_{\operatorname{#1}}}}
\newcommand{\N}[1]{\ensuremath{N_{\operatorname{#1}}}}
\newcommand{\fdNdp}[1]{\ensuremath{\mathrm{d}N_{\operatorname{\mu}}^{\operatorname{#1}}/\mathrm{d}p}}
\newcommand{\dNdp}[1]{\fdNdp{#1}}
\newcommand{\R}[1]{\ensuremath{R_{\operatorname{#1}}}}
\newcommand{\Tm}[1]{\ensuremath{T_{\operatorname{#1}}}}
\newcommand{\ltime}[1]{\ensuremath{\tau_{\operatorname{#1}}}}
\newcommand{\fnc}[1]{\ensuremath{f_{\operatorname{#1}}}}

\begin{document}

\title{First measurement of the low-momentum negative-muon beam for the muon-to-electron conversion experiment at J-PARC}

\author{%
\collaborator{The COMET Collaboration}
K.~Oishi\thanks{E-mail: kou.oishi@kek.jp}$^{10,12}$, 
R.~Abramishvili$^{8}$, 
G.~Adamov$^{8}$, 
J.~C.~Angelique$^{3}$, 
K.~Aoki$^{10}$, 
M.~Aoki$^{25}$, 
Y.~Arimoto$^{10}$, 
A.~N.~B.~Az$^{29}$, 
I.~Bagaturia$^{8}$, 
D.~Bauer$^{12}$
% ...
}

\affil{%
% ...
$^{3}$Université de Caen Normandie, ENSICAEN, CNRS/IN2P3, LPC Caen, Caen 14050, France\\%
% ...
$^{8}$Georgian Technical University (GTU), Tbilisi 0160, Georgia\\%
% ...
$^{10}$High Energy Accelerator Research Organization (KEK), Tsukuba, Ibaraki 305-0031, Japan\\%
% ...
$^{12}$Imperial College London, London SW7 2BW, UK\\%
% ...
$^{25}$The University of Osaka, Osaka 560-0043, Japan\\%
% ...
$^{29}$Universiti Teknologi Malaysia, Johor Bahru, Johor 81310, Malaysia\\%
% ...
%
}

\begin{abstract}
  The COMET \PhaseAlpha experiment, the initial beam-operation stage of the COMET muon-to-electron conversion programme at J-PARC, took place in 2023 and marked the first beam operation at the COMET facility.
  It served to commission the delivery of a bunched, slow-extracted \qty{8}{\GeV} proton beam and the muon beamline.
  Its objective was to demonstrate experimentally that low-momentum negative muons could be transported to the experimental area through the Transport Solenoid (TS).
  We report a measurement of the transported muon beam using the Range Counter, a dedicated detector that identifies stopped negative muons by detecting delayed electrons emitted after the muons stop in a copper absorber.
  From the measured decay-time spectra, we reconstruct the number of stopped muons, the corresponding momentum spectrum, and the transverse beam profile.
  The observation of the stopped-muon component demonstrates that slow negative muons were transported through the TS to the experimental area.
  The reconstructed momentum spectrum indicates that the transported beam is concentrated in the low-momentum region.
  The non-uniform transverse profile is qualitatively consistent with charge drift in the curved solenoidal magnetic field.
  A Geant4-based simulation reproduces the overall dependence on degrader thickness and detector position; the measured yields are systematically higher by roughly \qty{30}{\percent}.
  These results provide the first experimental benchmark of low-momentum negative-muon transport in COMET and establish the basis for future beam characterisation and production studies.
\end{abstract}

\subjectindex{C31, G12, H10}

\maketitle

\section{Introduction}
\label{sec:Intro}

The COMET experiment will search for charged lepton flavour violation through coherent muon-to-electron (\textmu-e) conversion ($\mu^- N \to e^- N$) at the Japan Proton Accelerator Research Complex (J-PARC)~\cite{COMET}.
The \textmu-e conversion search is planned in two stages, Phase-I and Phase-II, with expected single-event sensitivities of $\mathcal{O}(\num{e-15})$ and $\mathcal{O}(\num{e-17})$, respectively.
COMET requires the development of detectors and analysis methods for identifying the conversion signal, as well as a dedicated proton-beam operating mode and a muon transport system suited to the experiment.
The development of the beamlines is therefore a crucial part of the experiment.

In the \textmu-e conversion search, the signal consists of electrons from coherent conversion of negative muons stopped in the aluminium target, whereas much of the beam-related background is prompt relative to the proton arrival time.
For that reason, COMET adopts a pulsed proton beam and observes the signal electrons in a delayed time window after each pulse.
This time structure suppresses prompt backgrounds and retains sensitivity to the delayed \textmu-e conversion signal.
In addition, COMET uses a primary proton beam energy of \qty{8}{\GeV} to minimise antiproton production, because antiproton annihilation can produce electrons that mimic the signal.
These beam requirements are different from the nominal operating conditions of the J-PARC Main Ring (MR) accelerator.
A dedicated bunched slow-extraction scheme has therefore been developed for COMET to deliver an \qty{8}{\GeV} pulsed proton beam with the required time structure.
Early studies of the MR beam-operation performance have been reported~\cite{COMETBeamPoS402104}.

The Phase-I beamline consists of three superconducting solenoids.
The proton beam is injected into the graphite Pion Production Target, and secondary pions and muons generated in the backward direction are captured by the Pion Capture Solenoid (PCS) and guided through the curved Transport Solenoid (TS) system to the downstream Detector Solenoid (DS).
The PCS is energised at \qty{4.4}{\tesla} to maximise the capture of pions and muons.
Compared with those generated in the forward direction, the backward secondaries are expected to have a softer momentum spectrum, with a larger fraction in the low-momentum region relevant to the \textmu-e conversion search.
This is advantageous because higher-momentum particles are less likely to result in muons stopping in the aluminium target.
Instead, these particles tend to contribute to backgrounds.
The DS contains the aluminium target and the tracking detector system for the \textmu-e conversion search.
As of 2026, the full solenoid system has been installed and is being commissioned towards Phase-I physics data-taking.

The role of the TS is not merely to connect the production region to the detector region.
In a curved solenoidal field, charged particles experience a transverse drift whose direction depends on the charge sign and whose magnitude depends on the momentum.
The dipole field superimposed on the curved solenoid guides negative particles in the desired low-momentum range through the transport channel.
Particles with the wrong charge sign or unsuitable momentum are displaced from the central trajectory and are therefore more effectively removed by the beamline apertures and collimation system.
The performance of this transport is a central prerequisite for the Phase-I programme, because the conversion search depends on delivering a sufficiently intense and well-characterised slow-muon beam to the muon stopping target.

These beam and transport requirements motivated an initial beam operation dedicated to validating the proton beam delivery and the TS before the full Phase-I detector system and physics programme are realised.
Accordingly, the \PhaseAlpha experiment, the initial beam-operation stage of the COMET muon-to-electron conversion programme at J-PARC, was carried out in February and March 2023.
Its main goals were to commission proton-beam extraction to the COMET facility, operate the TS with beam, and establish direct experimental access to the slow negative-muon beam transported to the experimental area.
At the time of \PhaseAlpha, the PCS was still under development, but the delivery of the bunched, slow-extracted proton beam and the operation of the TS with beam still required commissioning.
The \PhaseAlpha configuration therefore operated the TS without the PCS magnetic-field system, providing a focused validation of the TS transport itself.
This was technically important because the transport of a muon beam through a curved solenoid had previously been demonstrated with the \qty{36}{\degree} curved solenoid of the MuSIC beamline~\cite{MUSIC}, whereas the COMET TS extends the transport bend to \qty{90}{\degree}.
The present measurement thus benchmarks low-momentum negative-muon transport through the TS in this TS-only configuration; Phase-I operation with the PCS corresponds to a different secondary-particle collection and upstream transport configuration.
For that reason, directly measuring the transported slow-muon beam in the experimental area was important in its own right: the beam data tested whether the TS provided the charge and momentum selection required for Phase-I.
Because it combined infrastructure commissioning with a first beam measurement, \PhaseAlpha occupies a special position in the COMET programme: it is both an engineering milestone and a physics validation dataset.

The transported beam was characterised through measurements of its momentum and transverse profile.
This paper reports these measurements, including validation studies and scans of the beam momentum and transverse position.
The results are compared with a Geant4-based simulation of the transport and detector response, which provides the first experimental benchmark of the COMET transport model.
\section{Experimental setup and datasets}
\label{sec:Exp}

\begin{figure}[tb]
  \centering
  \includegraphics[width=\linewidth]{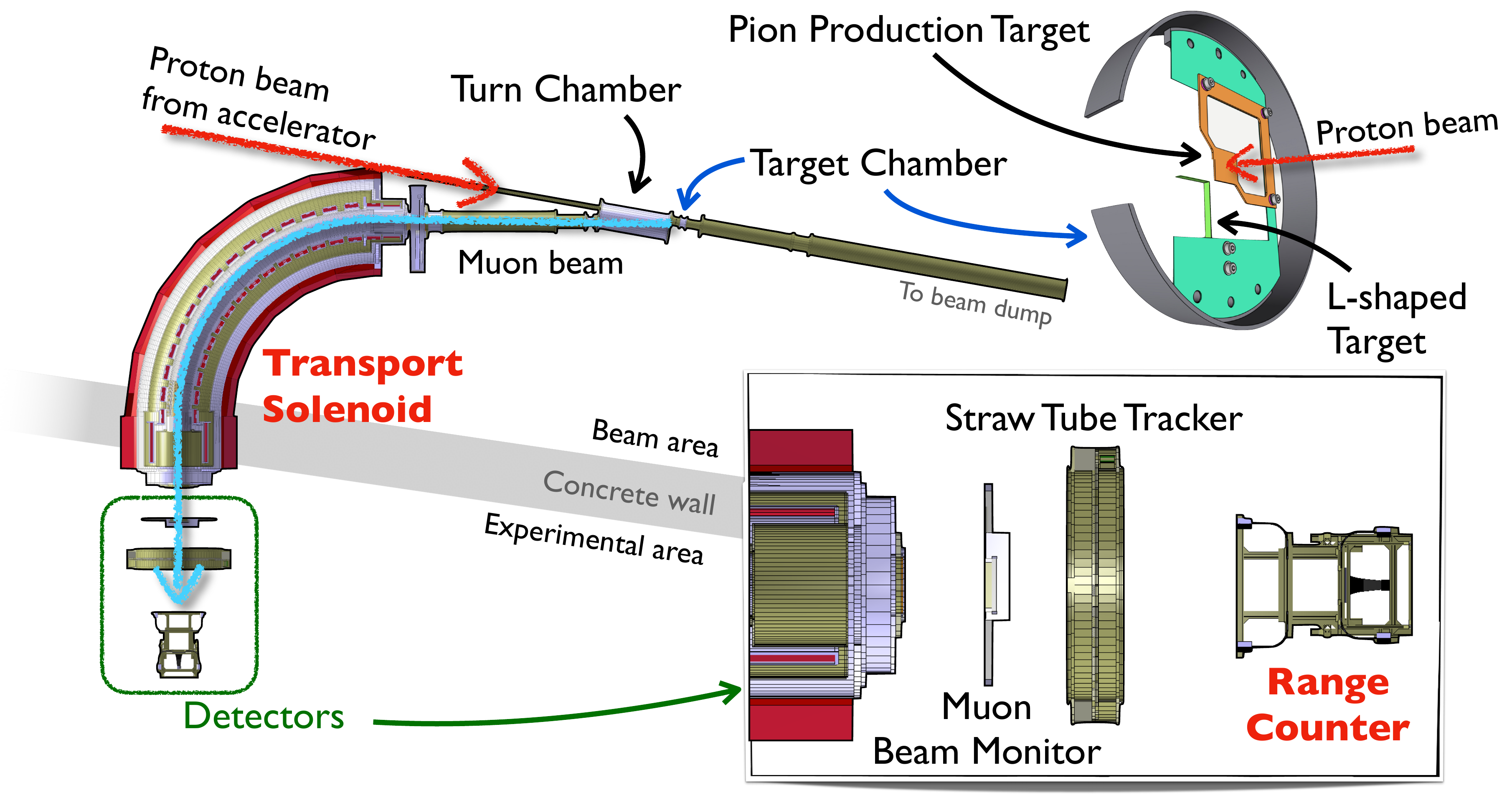}
  \caption{
    Schematic illustration of the \PhaseAlpha detector configuration and beamline geometry.
    The proton beam strikes the Production Target, and secondary particles produced in the backward direction enter the Transport Solenoid, which delivers slow negative muons to the experimental area.
    The Muon Beam Monitor, Straw Tube Tracker, and Range Counter (RC) were installed there; only the RC data are used in the analysis reported in this paper.
  }
  \label{fig:Exp.Beamline}
\end{figure}

The \PhaseAlpha experimental setup is illustrated schematically in \fig{Exp.Beamline}, which shows the proton beamline, TS, experimental area, and detector locations.
Three detectors were installed in the experimental area: the Muon Beam Monitor (MBM)~\cite{MBM}, the Straw Tube Tracker (STT)~\cite{StrawTrk}, and the Range Counter (RC)~\cite{RangeCounter}.
In the main beam period in March 2023, the RC, originally developed for measuring transported muons in the momentum range relevant to COMET~\cite{RangeCounter}, was used as the primary detector for beam characterisation.
The MBM is a scintillating fibre hodoscope located near the downstream end of the TS and was used to observe the charged beam profile and bunch-by-bunch intensity variations.
The STT consisted of one station of straw-tube proportional chambers and served as a demonstrator for tracking detectors in future COMET physics programmes.
The MBM and STT were used for beam commissioning and characterisation during \PhaseAlpha, whereas the physics results presented in this paper are derived exclusively from the RC data.

\subsection{Beamline components}
\label{sec:Exp.Beamline}

\begin{figure}[tb]
  \centering
  \includegraphics[width=\linewidth]{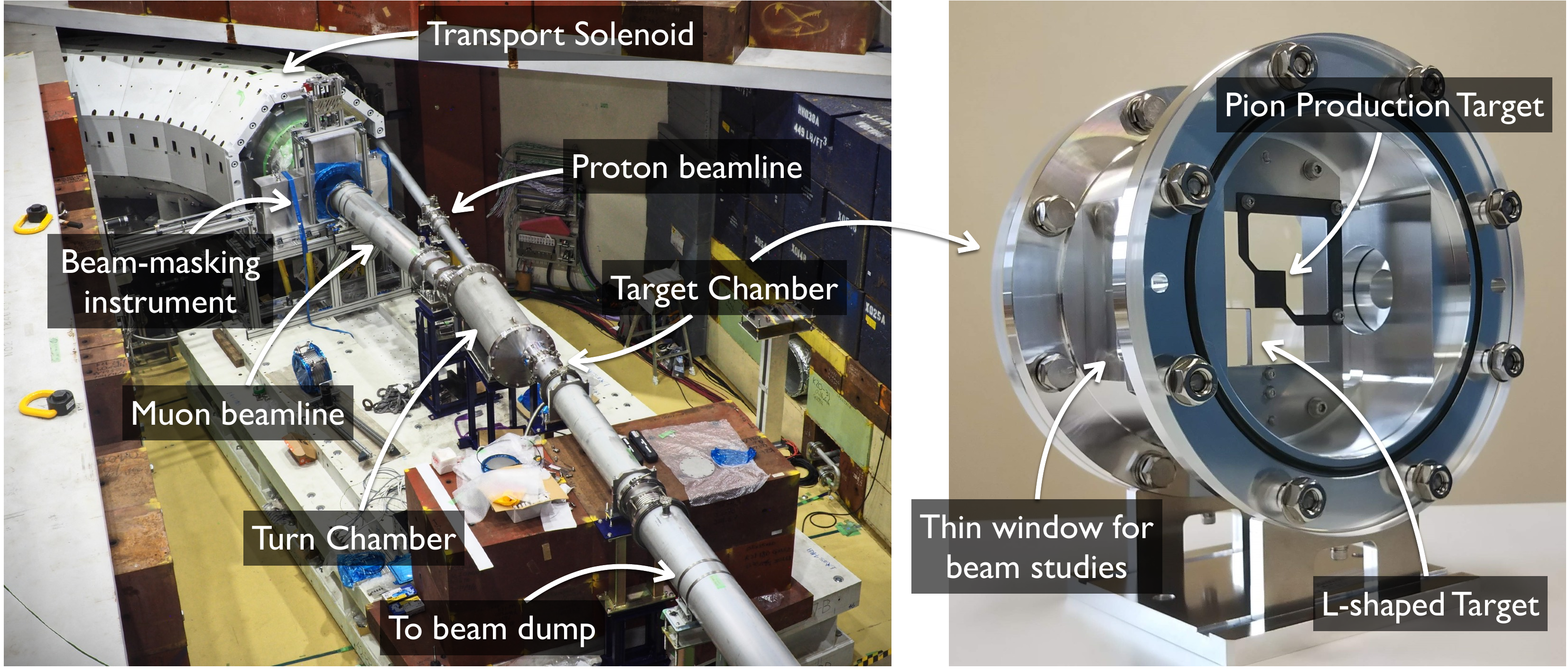}
  \caption{
    Photograph of the upstream region of the COMET beamline and the Target Chamber, corresponding to the schematic layout in \fig{Exp.Beamline}.
    The proton and muon beamlines meet at the Turn Chamber; downstream, the Target Chamber contains the Pion Production Target and L-shaped target, and the proton beamline continues to the beam dump.
    The Transport Solenoid and beam-masking instrument are also shown.
  }
  \label{fig:Exp.Comm.Setup}
\end{figure}

\fig{Exp.Comm.Setup} shows the proton and muon beamlines together with the Production Target and its supporting beam duct.
In the Turn Chamber at the centre of the photograph, the proton beamline connects to the secondary muon beamline.
A beam-masking instrument with removable slits was installed between the muon beamline and the TS for supplementary measurements of beam-transport dynamics, but was not used in the main measurement reported in this paper.

The Target Chamber is located downstream of the Turn Chamber and contains the Production Target.
The Production Target was a thin carbon-fibre-reinforced carbon (C/C) composite sheet with dimensions of \qtyproduct[product-units = power]{20 x 20 x 1.1}{\mm}~\cite{CX761}.
An L-shaped target next to the Production Target was also used for beam characterisation.
It was fabricated by bending a \qty{0.1}{\mm}-thick sheet of 304-grade austenitic stainless steel and had a length of \qty{4}{\mm} along the proton-beam direction.

The secondary muon beamline at the TS entrance is \qty{4}{\cm} lower than the proton beamline, so the Production Target is \qty{4}{\cm} above the muon-beamline axis in \PhaseAlpha.\footnote{For Phase-I, this geometric offset compensates the approximately \qty{4}{\cm} downward displacement of the proton beam in the Pion Capture Solenoid field.}
The TS entrance subtends a solid angle of approximately \qty{3.4}{\milli\steradian}, or \qty{0.027}{\percent} of $4\pi$, when viewed from the Production Target.

Secondary particles produced in the backward direction at the Production Target enter the TS.
For \PhaseAlpha, the TS was operated at \qty{1.5}{\tesla}\footnote{\qty{1.5}{\tesla} is half of the nominal Phase-I field strength for safety reasons, but is sufficient to retain the essential transport behaviour of slow negative particles in a curved solenoid under the initial beam conditions.} with a superimposed dipole field of \qty{50}{\mT}.

\subsection{Proton beam operation and commissioning}
\label{sec:Exp.Comm}

\PhaseAlpha demonstrated the first successful delivery of \qty{8}{\GeV} protons to the COMET beamline using a dedicated bunched slow-extraction scheme.
The spill cycle is \qty{9.6}{\second}, and the extraction period within each cycle is approximately \qty{0.5}{\second}.
The beam was transported to the Production Target region and remained stable throughout the data-taking period.

For COMET operation, four of the nine radio-frequency buckets in the MR are filled, and the resulting four proton bunches in one MR revolution are labelled K1--K4 in chronological order.
The bunch spacing is about \qty{1.2}{\micro\second}, except for the interval from K4 to K1, which is about \qty{1.8}{\micro\second}.
These bunch timings provide the time reference for the RC data acquisition.
The beam power in \PhaseAlpha was about \qty{0.23}{\kW}, reduced from the Phase-I value of \qty{3.2}{\kW} for radiation safety, corresponding to approximately \num{1.7e12} protons on target (POT) per spill.
The extraction efficiency from the MR was more than \qty{99}{\percent}, and the spill duty factor was about \qty{76}{\percent}~\cite{PAC36}.

During proton-beam tuning, the delivered beam was checked using a phosphor plate installed immediately upstream of the beamline dump.
The phosphor images confirmed that the beam spot reached the intended position after tuning.
The transverse beam size at the Production Target position was then measured using a stainless-steel L-shaped target.
For this measurement, the beam optics were adjusted to scan the beam across the horizontal and vertical edges of the fixed L-shaped target at the Production Target position.
When the target intercepted the beam, secondary particles scattered approximately perpendicular to the proton-beam direction were detected by requiring a coincidence among three plastic scintillators aligned as a telescope beside the Target Chamber.
From these horizontal and vertical edge scans, the Gaussian beam widths at the Production Target position were measured to be \qty{2.14}{\mm} and \qty{3.88}{\mm}, respectively~\cite{JPSCP45_011070}.

A compact proton beam monitor based on TiO$_2$ wide-bandgap semiconductor sensors~\cite{Chetry23TiO2} also recorded proton signals as an auxiliary diagnostic for beam operation.

\subsection{Range Counter}
\label{sec:Exp.RC}

\begin{figure}[tb]
  \centering
  \includegraphics[width=\linewidth]{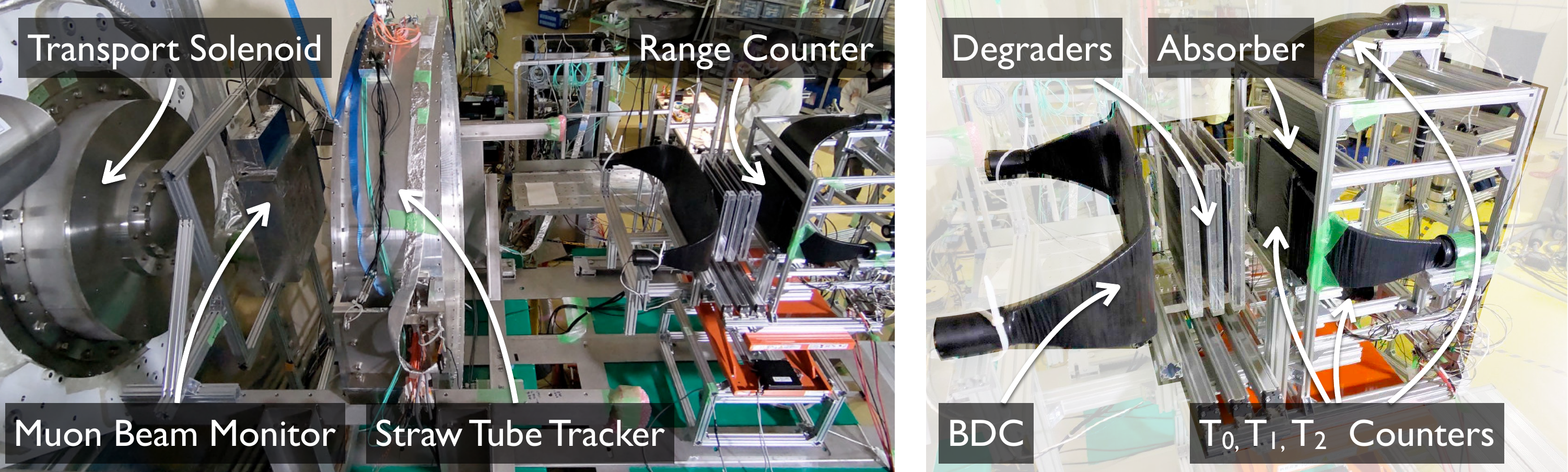}
  \caption{
    Photographs of the detector system used in the \PhaseAlpha beam measurement.
    The left panel shows the detector system installed in the experimental area, and the right panel shows a detailed view of the Range Counter.
    The copper absorber is inserted at the location indicated by the arrow between the \T0 and \T1 Counters.
  }
  \label{fig:Exp.Detectors}
\end{figure}

\begin{figure}[tb]
  \centering
  \includegraphics[width=\linewidth]{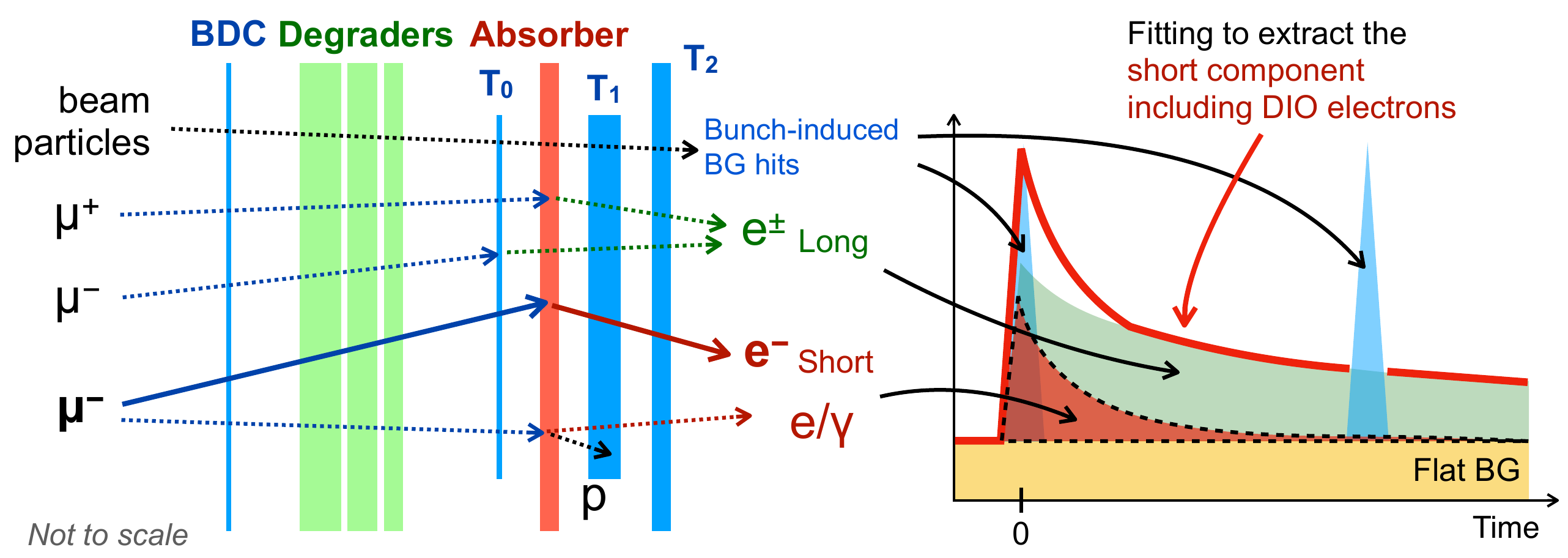}
  \caption{
    Schematic of the negative-muon measurement with the Range Counter.
    The blue components are the BDC and the \T0, \T1, and \T2 plastic-scintillator counters; the graphite degraders and copper absorber are green and red, respectively.
    An incoming negative muon is tagged upstream, slowed by the degraders, and stopped in the copper absorber.
    Delayed electrons from muon decay are detected downstream to identify stopped negative muons.
    The corresponding decay-time distribution contains a short-lifetime component from electrons emitted by muon decay in the copper absorber and a longer-lived component from muons stopping in lighter counter materials.
    It also contains a flat accidental component and narrow bunch-induced peaks produced by particles arriving at the accelerator bunch timings.
  }
  \label{fig:Exp.RC}
\end{figure}

The RC, displayed in \fig{Exp.Detectors} with the other detectors, is the key detector for the beam characterisation measurement~\cite{RangeCounter}.
Its method for measuring negative muons is illustrated schematically in \fig{Exp.RC}.
As shown in the figure, the RC comprises four plastic scintillator timing counters: the upstream Beam-defining Counter (BDC) and \T0 Counter, and the downstream \T1 and \T2 Counters.

The BDC--\T0 coincidence tags a charged particle transported from the TS; graphite degraders then reduce its momentum, allowing a negative muon to stop in the copper absorber.
The downstream \T1 and \T2 Counters detect electrons from muon decay in orbit (DIO), providing the signature of stopped negative muons.
An RC signal candidate is therefore defined by an upstream BDC--\T0 coincidence followed by a \T1--\T2 coincidence from a delayed electron with energy deposits above the selection thresholds.
For hit times $T_0$, $T_1$, and $T_2$ in the \T0, \T1, and \T2 Counters, respectively, the decay-time variable is defined as $\Tm{decay} = (T_1+T_2)/2-T_0$.\footnote{Here and below, upright \T0, \T1, and \T2 denote the counters, while italic $T_0$, $T_1$, and $T_2$ denote the corresponding hit times.}

Particles arriving at the accelerator bunch timings can also produce coincident hits in the downstream counters, creating narrow bunch-induced peaks at the corresponding values of $\Tm{decay}$.
Their spacing reflects the accelerator bunch structure.
For stopped negative muons, DIO electrons outside the bunch-induced peak regions form the signal component: the short-lifetime component is attributed to muonic copper atoms formed in the absorber, while a longer component arises from stops in lighter RC materials.

The dimensions and readout of the RC components are as follows.
The upstream BDC and \T0 Counters are thin plastic scintillator plates with dimensions of \qtyproduct[product-units = power]{300 x 300 x 0.5}{\mm} and \qtyproduct[product-units = power]{200 x 200 x 0.5}{\mm}, respectively.
Both counters are read out from two sides and are used to tag an incoming charged particle from the TS.

The copper absorber, with dimensions of \qtyproduct[product-units = power]{300 x 300 x 0.5}{\mm}, was installed immediately behind the \T0 Counter.
It was followed by the downstream \T1 and \T2 Counters, which detected electrons emitted after a muon stopped in the copper absorber.
The \T1 and \T2 Counters have dimensions of \qtyproduct[product-units = power]{200 x 200 x 10}{\mm} and \qtyproduct[product-units = power]{300 x 300 x 5}{\mm}, respectively, and are read out from one side.

Between the BDC and the \T0 Counter, graphite degrader plates can be inserted remotely.
The total degrader thickness controls the incoming muon momentum band that can stop in the copper absorber, i.e., the effective momentum acceptance of the RC for the measurement.

\subsection{Trigger}
\label{sec:Exp.Trigger}

The data-acquisition (DAQ) trigger was
\begin{equation}
  \mathrm{BDC}\ \&\ \T0\ \&\ \overline{(\T1\ |\ \T2)}\ \&\ \mathrm{G_{spill}},
  \label{eq:Exp.Trigger.Main}
\end{equation}
where $\mathrm{G_{spill}}$ is a \qty{650}{\milli\second} DAQ gate timed to contain the full beam-extraction period with margin.
The BDC--\T0 coincidence tags a charged particle entering the RC, while the \T1 or \T2 veto suppresses particles penetrating through the absorber region at the trigger time.
A less restrictive trigger, $\mathrm{BDC}\ \&\ \T0\ \&\ \mathrm{G_{spill}}$, was used in selected validation data.
In addition, a \qty{2}{\milli\second} self-veto was applied after each accepted trigger to keep the DAQ rate within the stable operating range and to avoid event-building mismatches.
This self-veto is included in the DAQ inefficiency.

\subsection{Data acquisition and datasets}
\label{sec:Exp.Data}

Waveforms from the RC counters and timing-logic signals were recorded with digitisers.
The RC waveforms were acquired using a PicoScope 6824E at a sampling interval of \qty{0.8}{\nano\second} (\qty{1.25}{Gs\per\second}).
The waveform window was \qty{10}{\micro\second}, extending from \qty{-2}{\micro\second} to \qty{8}{\micro\second} relative to the trigger time.
A scaler module was also used to record the raw and accepted trigger counts for the evaluation of the DAQ efficiency.

The analysis presented here is based on datasets collected during the March 2023 beam time, organised into three groups.
The first group is the degrader scan, parameterised by the total graphite degrader thickness $d_{\rm deg}$, with datasets at $d_{\rm deg}=\qtylist{0;4;12;16;20;32}{\mm}$.
The second group is the beam-position scan, in which the RC was translated to several positions without a degrader.
The third group consists of validation measurements.

In the beam-position scan, the RC was placed at six positions labelled vertical.horizontal.
The labels combine three vertical positions (upper, centre, and bottom) with two horizontal positions (left and centre), as viewed from the downstream side; adjacent positions along either coordinate were separated by \qty{25}{\cm}.
Because COMET requires a low-momentum muon beam, the beam-position scan was carried out without a degrader, corresponding to the lower-momentum, highest-statistics configuration with an effective momentum band centred around \qty{37}{\MeVc}.
The beam-position scan therefore probes the spatial distribution of the beam in this COMET-relevant configuration.
These six no-degrader datasets, together with the five nominal-RC-position degrader-scan datasets, form the 11 datasets used in the decay-time fit in the main analysis.

The validation measurements examined how the measured decay-time spectrum depends on the TS magnetic field and the presence of the copper absorber.
For the TS-field comparison, data were taken with the TS magnet switched on and off; in the magnet-off run, the RC absorber and nominal trigger configuration were unchanged.
For the absorber comparison, data with the absorber installed and removed were taken with the less restrictive BDC--\T0 trigger to compare the delayed-electron yield between the two configurations while avoiding unnecessary veto bias in the absorber-removed configuration.
For both comparisons, each decay-time spectrum was normalised by the corresponding number of protons on target and DAQ efficiency.
\section{Experimental data analysis}
\label{sec:Analysis}

\subsection{Event reconstruction and decay-time spectrum}
\label{sec:Analysis.Signal}

The analysis begins with the digitised waveforms recorded for each counter in every triggered event.
A threshold-based waveform search identifies individual pulses and extracts their times and amplitudes.
For each triggered event, the BDC--\T0 coincidence is reconstructed within a \qty{50}{\nano\second}-wide trigger-timing window in the recorded waveform, using the same coincidence logic as in \eq{Exp.Trigger.Main}.
This coincidence defines the event time reference, $T_0$.
The timing coincidence criteria were optimised using the observed timing distributions of the relevant counters.

For each accepted trigger, the waveform data are then searched offline for coincident \T1--\T2 hit pairs with energy deposits above the thresholds, giving the hit times $T_1$ and $T_2$ and the signal candidates defined in \sect{Exp.RC}.
Because multiple \T1--\T2 hit-pair candidates can satisfy the selection, a single event can contribute more than one $\Tm{decay}$ entry.
The selected candidates form the reconstructed $\Tm{decay}$ distributions.
The \T1 or \T2 veto in the hardware trigger prevents candidates from being accepted around $\Tm{decay}=0$.

\subsection{Decay-time fit}
\label{sec:Analysis.Fit}

The short-lifetime, long-lifetime, and flat component yields are extracted with an extended maximum-likelihood fit.
The flat component accounts for random hit combinations and delayed background activity.
Each of the 11 main scan datasets is divided into four spectra labelled K1--K4; the resulting $11\times4=44$ spectra are fitted simultaneously.

\begin{figure}[tb]
\centering
\begin{minipage}[t][0.31\textheight][c]{0.49\linewidth}
  \centering
  \includegraphics[width=\linewidth,height=0.26\textheight,keepaspectratio]{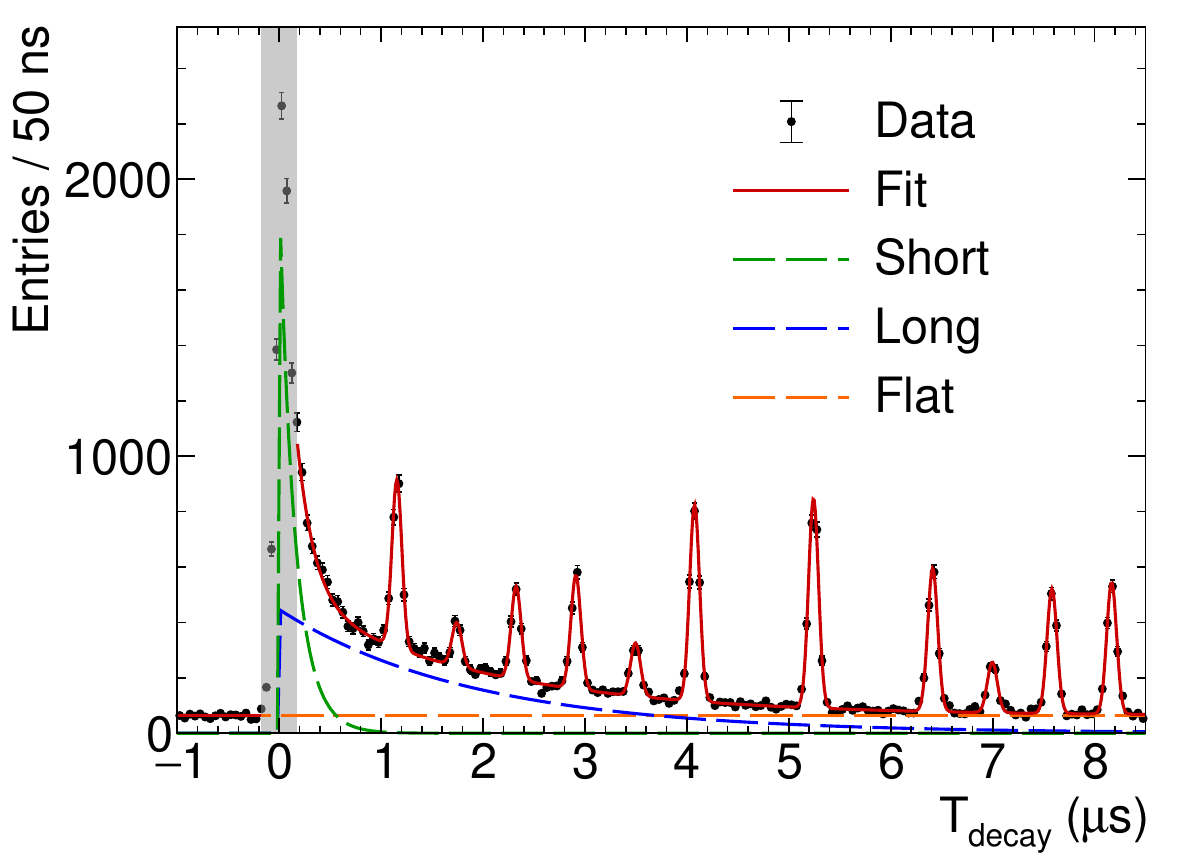}
  \subcaption{Merged prefit of all datasets}
\end{minipage}
\hfill
\begin{minipage}[t][0.31\textheight][c]{0.49\linewidth}
  \centering
  \includegraphics[width=\linewidth,height=0.26\textheight,keepaspectratio]{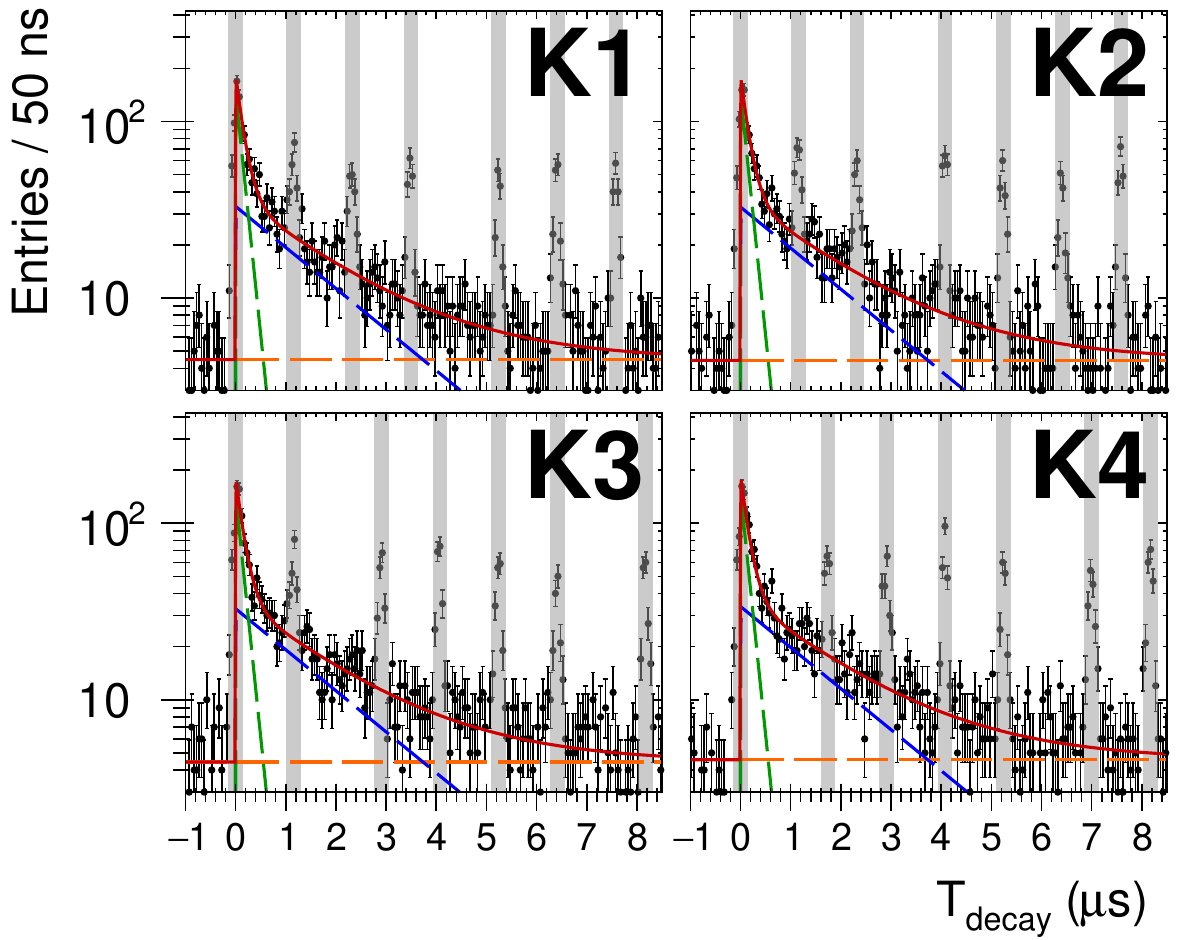}
  \subcaption{Simultaneous-fit subset}
\end{minipage}
\caption{
  Decay-time distributions used to define the bunch-induced-peak masks and extract the component yields.
  (a) The distribution obtained by merging all datasets. The red curve is the auxiliary prefit used to determine the positions and widths of the masks.
  (b) The four K1--K4 spectra from the no-degrader dataset at the central position, shown as a representative subset of the result from the simultaneous fit to all 44 spectra.
  The simultaneous fit separates the short- and long-lifetime exponential components from the flat accidental component.
  The grey shaded regions indicate the corresponding masked time regions.
}
\label{fig:Analysis.PromptPeaks}
\end{figure}

Before the simultaneous fit, the bunch-induced peaks are identified in the merged distribution with an auxiliary prefit.
As shown in \fig{Analysis.PromptPeaks}(a), this prefit determines their positions and common width.
The bunch-induced peak positions depend on the bunch label, so each spectrum uses the corresponding masked time regions shown in \fig{Analysis.PromptPeaks}.
These regions are excluded from the simultaneous fit.
\fig{Analysis.PromptPeaks}(b) displays only the four K1--K4 spectra from the no-degrader dataset at the central RC position, as a representative subset of the simultaneous-fit result for all 44 spectra.
The short and long lifetimes, \ltime{short} and \ltime{long}, are common to all 44 spectra.
For each dataset, \N{short}, \N{long}, and \N{flat} are the fitted total yields of the short-lifetime, long-lifetime, and flat components, respectively.
Fitted relative bunch populations distribute each total among K1--K4.
The reported component yields are the totals over K1--K4 for each dataset.

The fit systematic uncertainty is defined from stability checks in which the bunch-induced peak mask regions and the initial fit parameters are varied.
The root-mean-square (RMS) variation of the fitted yield in these checks is assigned as the fit systematic uncertainty.

As a validation of the fit procedure, the fitted lifetimes are
\begin{align}
  \ltime{short} &=
  0.1694^{+0.0122}_{-0.0117}\,(\mathrm{stat.})^{+0.0002}_{-0.0009}\,(\mathrm{sys.})~\unit{\micro\second}, \\
  \ltime{long} &=
  1.9082^{+0.0615}_{-0.0593}\,(\mathrm{stat.})^{+0.0004}_{-0.0033}\,(\mathrm{sys.})~\unit{\micro\second}.
\end{align}
\ltime{short} agrees with the reported lifetime of negative muons stopped in copper, $(0.1635\pm0.0010)~\unit{\micro\second}$~\cite{MuonicAtomLife.1}.
\ltime{long} is interpreted as an effective lifetime for muons stopping in lighter materials in the RC, with contributions from graphite degraders, plastic scintillators, and the aluminium frame, together with detector and time-selection effects.
It therefore represents a mixture of stopping locations rather than a single-material exponential and may be shorter than the measured negative-muon lifetime in carbon, $(2.0263\pm0.0015)~\unit{\micro\second}$~\cite{MuonicAtomLife.1}.
\subsection{Reconstruction of the stopped-muon yield}
\label{sec:Analysis.Nstop}

The stopped-muon yield per proton on target (POT) is reconstructed from the fitted short-component yield, \N{short}, as
\begin{equation}
  \N{stop} =
  \frac{1}{\N{p}\,\eff{ext}\,\eff{DAQ}}
  \frac{\N{short}\,\pur{DIO}}
  {\eff{trig}\,\fnc{trig}\,(1-\eff{veto})\,\R{DIO}\,\accep{DIO}}.
  \label{eq:Analysis.Nstop}
\end{equation}
The first factor gives the beam and DAQ normalisation, while the second contains the RC response and selection corrections.
\N{p} is obtained by spill-by-spill integration of the proton intensity measured in the MR over the corresponding data-taking period.
For runs that started or ended within a spill, the integrated intensity is weighted by the active data-taking fraction determined from event time stamps; the statistical uncertainty on \N{p} is negligible.
$\eff{ext}$ is the extraction efficiency of the proton beam, and hence $\N{p}\eff{ext}$ corresponds to POT.
The reported extraction efficiency exceeds \qty{99}{\percent}~\cite{PAC36}; we conservatively use $\eff{ext}=0.99^{+0.01}_{-0.00}$, with an upper uncertainty extending to the physical limit of unity.

The data acquisition efficiency, \eff{DAQ}, is evaluated as $N_{\rm rec}/N_{\rm raw}$.
Here, $N_{\rm raw}$ is the number of raw trigger signals counted by the scaler and $N_{\rm rec}$ is the number of events recorded by the DAQ.
As a consistency check, the scaler also counted the triggers accepted after the DAQ busy and self-veto logic, denoted here by $N_{\rm acc}$.
The ratio $N_{\rm rec}/N_{\rm acc}$ should be unity; its RMS spread over the main datasets is \qty{1.47}{\percent}.
This RMS is assigned as the relative systematic uncertainty on \eff{DAQ} to cover possible scaler or trigger-logic counting mismatches.
The statistical uncertainty on \eff{DAQ} is negligible.

The factor $\eff{trig}$ is the probability that a slow muon entering the RC and stopping in the absorber satisfies the upstream trigger condition in \eq{Exp.Trigger.Main}.
In \PhaseAlpha, each BDC and \T0 hit is formed from a left--right coincidence of the two readout sides, and the hardware trigger then requires a BDC--\T0 coincidence.
\eff{trig} is derived from the single-counter efficiencies measured in the RC performance study~\cite{RangeCounter}.
That study parameterises the efficiency as a function of the discriminator threshold normalised to the most probable waveform amplitude for minimum-ionising particles.
For the \PhaseAlpha operating thresholds, the BDC and \T0 efficiencies are recalculated from this parameterisation and then combined according to the hardware BDC--\T0 coincidence logic.
The BDC is upstream of the degrader. Therefore, \eff{trig} depends on the target momentum setting: the stopping-muon momentum determines the energy deposited in the BDC and hence the probability of passing its threshold.
The uncertainty on \eff{trig} includes the counter-efficiency fit uncertainty, the spread among beam positions in the reference measurement, and the beam-position spread propagated to the momentum-dependent efficiency.

Scaler data and waveform-amplitude checks identified one data-taking period with reduced BDC--\T0 trigger formation efficiency.
The factor $\fnc{trig}$ corrects this loss. After correcting for \eff{DAQ}, it is obtained by comparing the beam-normalised yield of triggered stopped-muon events in the affected period with that in stable periods.
The comparison gives a \qty{10.0}{\percent} reduction, so $\fnc{trig}=0.90$ is applied to the affected fraction of the data.
A \qty{2.9}{\percent} systematic uncertainty, assuming a uniform distribution, is assigned. Both \fnc{trig} and its uncertainty are scaled by the affected fraction of each dataset.

$(1-\eff{veto})$ corrects for valid events lost through accidental veto activity.
The BDC--\T0-triggering muons are synchronised with the accelerator bunch timing, while unrelated charged particles can hit either the \T1 or the \T2 Counter in the veto window.
\eff{veto} is evaluated with waveform data from pre-trigger bunches, in which \T1 or \T2 hits are uncorrelated with the trigger.
The \T1 and \T2 veto thresholds used in data taking are applied to these waveforms to measure the pre-trigger hit probability.
The accidental-veto probability is calculated as the product of the probability for a bunch-synchronised \T1 or \T2 hit and the probability that the hit falls within the effective veto window.
This gives $\eff{veto}=(0.36\pm0.02)\%$, or equivalently the common survival factor $(1-\eff{veto})=0.9964\pm0.0002$.

The DIO probability $\R{DIO}$ is the branching probability for DIO in a stopped muonic copper atom.
$\R{DIO}$ is taken as the uncertainty-weighted average of the published DIO probabilities for muonic copper reported in~\cite{MuonicAtomLife.1}, giving $\R{DIO}=\num{0.0720}\pm\num{0.0004}$.

$\pur{DIO}$ is the fraction of selected short-component DIO electron candidates that are true DIO electrons from muons stopped in the copper absorber.
$\accep{DIO}$ is the acceptance for such true DIO electrons in the \PhaseAlpha geometry with the same \T1--\T2 coincidence and deposited-energy selection as used in data.
Their definitions and general evaluation method follow the RC design study~\cite{RangeCounter}.
For the \PhaseAlpha analysis, they were re-evaluated with the \PhaseAlpha geometry, measured RC alignment, finite counter gaps, and RC-position-dependent stopped-muon distributions.
The evaluation uses simulated DIO events and events induced by muon capture. The parent muon is required to traverse both the BDC and the \T0 Counter and stop in the absorber. The same \T1--\T2 coincidence and energy-threshold selection as in data is then applied.
The deposited-energy threshold was determined from the dependence of \pur{DIO} and \accep{DIO} on that requirement.
The adopted threshold, 40\% of the most probable signal size in the downstream counters, improves DIO purity while keeping the acceptance stable against detector-resolution effects.
The resulting values are $\pur{DIO}\approx0.669$ and $\accep{DIO}\approx0.411$, each with an uncertainty of about 0.001 and only weak dependence on the RC position.

\tab{Analysis.ReconParameters} lists the parameters in \eq{Analysis.Nstop} that vary with the dataset or RC configuration.
The remaining inputs common to all configurations, $\eff{ext}$, $(1-\eff{veto})$, and $\R{DIO}$, are given above and treated as common factors in the reconstruction.
The uncertainty on \N{p} is negligible compared with the other reconstruction parameters.

Only the uncertainty from the likelihood fit to \N{short} is treated as the statistical component of \N{stop}.
The systematic component includes the fit stability term and the uncertainties of the reconstruction parameters.

\begin{table}[tb]
  \centering
  \caption{
    Dataset-dependent parameters used to obtain the stopped-muon yield \N{stop} from the fitted short-component yield \N{short} in \eq{Analysis.Nstop}.
    \N{p} is the integrated number of protons measured in the Main Ring, \eff{DAQ} is the data-acquisition efficiency, \eff{trig} is the BDC--\T0 trigger efficiency, and \fnc{trig} corrects for reduced trigger-formation efficiency during one data-taking period.
    \pur{DIO} and \accep{DIO} are the purity and acceptance, respectively, for decay-in-orbit electrons from muons stopped in the copper absorber.
    The combination of the first two columns identifies each RC configuration.
    The first column gives $d_{\rm deg}$, the total graphite degrader thickness.
    The RC positions are denoted by labels of the form vertical.horizontal, with vertical = upper, centre, or bottom and horizontal = left or centre.
    Parentheses denote the uncertainty in the last quoted digits where applicable.
    The uncertainty on \N{p} is negligible.
  }
  \label{tab:Analysis.ReconParameters}
  \sisetup{separate-uncertainty = false}
  \setlength{\tabcolsep}{5.5pt}
  \renewcommand{\arraystretch}{1.22}
  \begin{tabular}{
    S[table-format=2.0]
    l
    S[table-format=4.1]
    c
    c
    c
    c
    c
  }
    \toprule
    {$d_{\rm deg}$} & {RC position} & {$\N{p}$} & {$\eff{DAQ}$} & {$\eff{trig}$} & {$\fnc{trig}$} & {$\pur{DIO}$} & {$\accep{DIO}$} \\
    {(\unit{\mm})} & {} & {($10^{12}$)} & {} & {} & {} & {} & {} \\
    \midrule
     0 & bottom.left   & 1820.2 & \num{0.560(8)} & \num{0.922(66)} & \num{1.000}     & \num{0.668(1)} & \num{0.411(1)} \\
     0 & bottom.centre & 1728.9 & \num{0.543(8)} & \num{0.922(66)} & \num{1.000}     & \num{0.668(1)} & \num{0.410(1)} \\
     0 & centre.left   & 1815.4 & \num{0.552(8)} & \num{0.922(66)} & \num{1.000}     & \num{0.667(1)} & \num{0.409(1)} \\
     0 & centre.centre & 6225.3 & \num{0.531(8)} & \num{0.922(66)} & \num{1.000}     & \num{0.669(1)} & \num{0.411(1)} \\
     0 & upper.left    & 1663.9 & \num{0.608(9)} & \num{0.922(66)} & \num{1.000}     & \num{0.668(1)} & \num{0.414(1)} \\
     0 & upper.centre  & 1638.4 & \num{0.596(9)} & \num{0.922(66)} & \num{1.000}     & \num{0.668(1)} & \num{0.415(1)} \\
     4 & centre.centre & 5183.2 & \num{0.544(8)} & \num{0.868(74)} & \num{0.900(29)} & \num{0.669(1)} & \num{0.411(1)} \\
    12 & centre.centre & 3563.5 & \num{0.645(9)} & \num{0.797(72)} & \num{0.941(17)} & \num{0.669(1)} & \num{0.411(1)} \\
    16 & centre.centre & 7534.8 & \num{0.675(10)} & \num{0.767(69)} & \num{0.920(23)} & \num{0.669(1)} & \num{0.411(1)} \\
    20 & centre.centre & 6474.7 & \num{0.768(11)} & \num{0.740(66)} & \num{0.974(8)}  & \num{0.669(1)} & \num{0.411(1)} \\
    32 & centre.centre & 1148.7 & \num{0.801(12)} & \num{0.668(60)} & \num{1.000}     & \num{0.669(1)} & \num{0.411(1)} \\
    \bottomrule
  \end{tabular}
\end{table}

\subsection{Conversion to a momentum spectrum}
\label{sec:Analysis.Spectrum}

For the degrader scan, \N{stop} is converted into the differential momentum spectrum in the incident momentum domain as
\begin{equation}
  \frac{\mathrm{d}N_{\operatorname{\mu}}}{\mathrm{d}p} =
  \frac{\N{stop}}{\R{stop}\,\Delta p},
  \label{eq:Analysis.Nmu}
\end{equation}
where \R{stop} is the stopping fraction and $\Delta p$ is the effective momentum bin width.
This conversion provides a reference mapping from each degrader thickness to a representative incident-momentum interval, allowing the scan to be expressed as an effective momentum-spectrum estimate.
The mapping is evaluated with a dedicated RC stopping simulation, independent of detailed transported-beam dynamics, and its parameters are listed in \tab{Analysis.RangeCurve}.
For each degrader thickness, negative muons with a flat momentum distribution are generated uniformly across the BDC surface at a position upstream of the BDC and then propagated through the RC.
The stopping-efficiency distribution is the simulated probability for a muon incident at the BDC plane to stop in the absorber, evaluated as a function of its incident momentum.
Its mean defines the stopping momentum.
Because the simulated response is sufficiently symmetric, although non-Gaussian, its full width at half maximum (FWHM) is conservatively adopted as $\Delta p$, the finite momentum interval represented by that degrader configuration.
The stopping fraction \R{stop} is taken as the average stopping efficiency over that FWHM interval.

\begin{table}[tb]
  \centering
  \caption[Range curve parameters used in the momentum spectrum conversion.]{
    Range curve parameters used to convert the measured stopped-muon yield $\N{stop}$ into the momentum-spectrum estimate $\fdNdp{}$.
    The first column gives $d_{\rm deg}$, the total graphite degrader thickness.
    The stopping momentum is the mean of the simulated stopping-efficiency distribution as a function of the incident muon momentum at the BDC plane, and $\Delta p$ is its full width at half maximum (FWHM).
    The stopping fraction $\R{stop}$ is the average stopping efficiency over the FWHM interval.
  }
  \label{tab:Analysis.RangeCurve}
  \setlength{\tabcolsep}{6pt}
  \renewcommand{\arraystretch}{1.22}
  \begin{tabular}{
    S[table-format=2.0]
    S[table-format=2.1]
    S[table-format=2.1]
    S[table-format=1.1]
  }
    \toprule
    {$d_{\rm deg}$} & {Stopping momentum} & {$\R{stop}$} & {$\Delta p$} \\
    {(\unit{\mm})} & {(\unit{\MeVc})} & {(\unit{\percent})} & {(\unit{\MeVc})} \\
    \midrule
     0 & 37.1 & 96.9 & 9.9 \\
     4 & 51.8 & 91.1 & 4.6 \\
    12 & 67.0 & 79.5 & 3.0 \\
    16 & 72.4 & 74.4 & 2.9 \\
    20 & 77.2 & 69.7 & 2.9 \\
    32 & 88.9 & 61.2 & 3.3 \\
    \bottomrule
  \end{tabular}
\end{table}
\section{Simulation}
\label{sec:Sim}

The reconstructed observables are compared with a full Geant4 simulation of secondary particle production and transport through the COMET beamline~\cite{Geant4a,Geant4b,Geant4c}.
The Geant4 version and physics list used to estimate the negative-muon yield under the \PhaseAlpha beam conditions are 10.6.p3 and QGSP\_BERT~\cite{WrightKelsey15,Heikkinen03}, respectively.
The comparison presented in this paper therefore provides an experimental benchmark of this simulation.
The \PhaseAlpha geometry is fully implemented, including the facility building, beamline components, and detector assemblies.
The TS magnetic field map used for tracking was calculated numerically.
The field description has been validated against in situ Hall-probe measurements of the superconducting Transport Solenoid~\cite{Arimoto24TSField}.

\subsection{Production and transport model}
\label{sec:Sim.Model}

The simulation begins by injecting the primary proton beam, with an energy of \qty{8}{\GeV}, into the Production Target.
Particle production is simulated in two statistically independent datasets: the target-induced and non-target-induced datasets.
These datasets represent non-overlapping parts of the transport calculation.
In the target-induced dataset, secondary particles generated inside the Production Target are transported downstream, while primary protons are terminated after leaving it.
In the non-target-induced dataset, the phase-space information of protons transmitted through the Production Target is recorded.
This information provides the input to a second simulation step, in which secondary particles produced in downstream materials are simulated.

This scheme separates the contributions from the Production Target and from beamline components surrounding and downstream of it.
The two datasets are therefore evaluated separately and summed when comparing with the data.

Particles transported through the TS to the detector region provide common inputs to the final simulation step, allowing the different RC positions and degrader configurations to be evaluated with the same upstream beam sample.

\subsection{Simulated observables and uncertainties}
\label{sec:Sim.Observables}

For each RC configuration, the simulated yield is calculated by counting negative muons that penetrate both the BDC and the \T0 Counter and stop in the copper absorber.
This definition corresponds directly to the reconstructed \N{stop} obtained from data.
The conversion in \eq{Analysis.Nmu} is then applied so that the simulation is expressed in the same effective observable as the data-derived momentum spectrum, including the same range-curve mapping and finite momentum acceptance.

The sensitivity to material assumptions is also evaluated by changing two geometry parameters.
The thickness of the MBM light-shielding window, which is made of aluminium foil, is varied over \qtylist{100;150;200}{\micro\meter} to cover a possible manufacturing ambiguity.
The graphite degrader, a major source of material along the beamline, is modelled with its thickness scaled by \qtylist{97;100;103}{\percent}.
For each RC configuration, the mean over these variations is used as the simulated central value, and the RMS of the variation is quoted as a geometry/material-model variation.
The dominant uncertainty is statistical because few simulated events contain muons that satisfy the stopping definition, and the high computational cost of simulating pion-producing interactions of primary protons limits the sample size.
The geometry/material-model variation is smaller than this statistical uncertainty.
\section{Experimental results}
\label{sec:Results}

\subsection{Validation measurements}
\label{sec:Results.Validation}

\begin{figure}[tb]
\centering
\begin{minipage}[t]{0.48\linewidth}
  \centering
  \includegraphics[width=\linewidth]{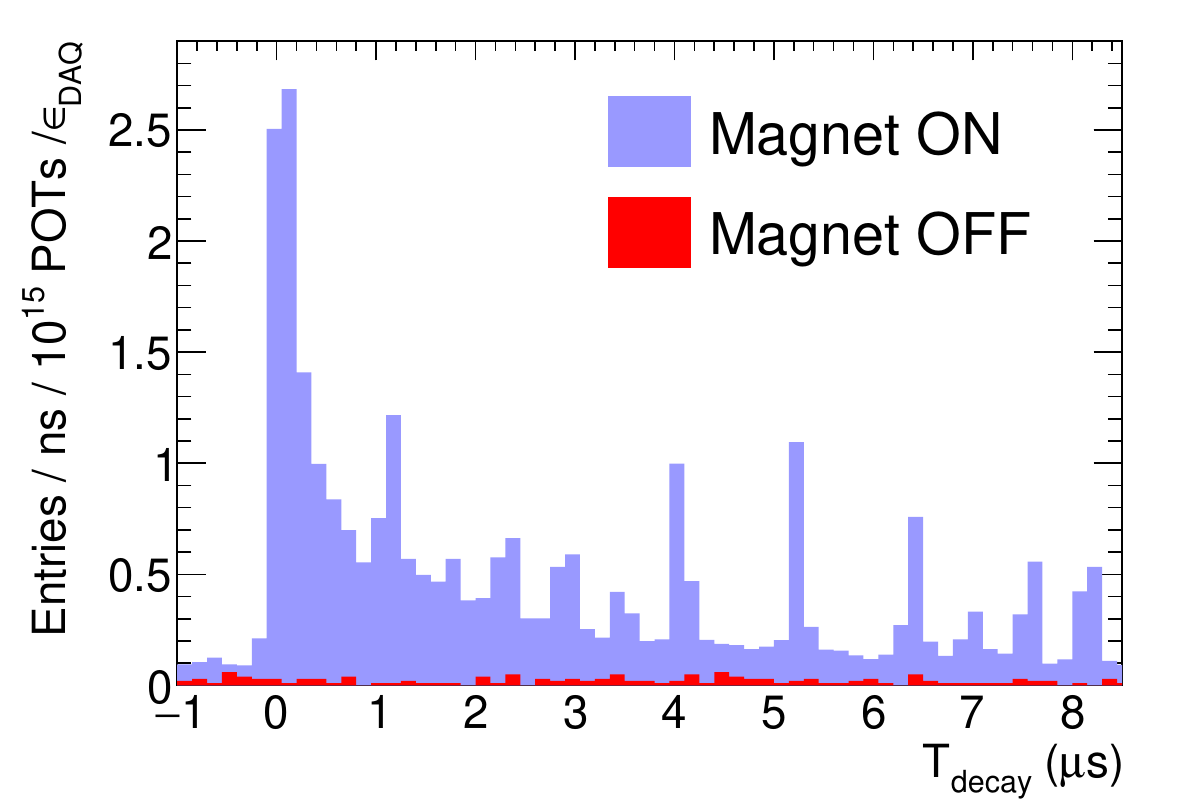}
  \subcaption{Effect of the TS magnetic field}
\end{minipage}
\hfill
\begin{minipage}[t]{0.48\linewidth}
  \centering
  \includegraphics[width=\linewidth]{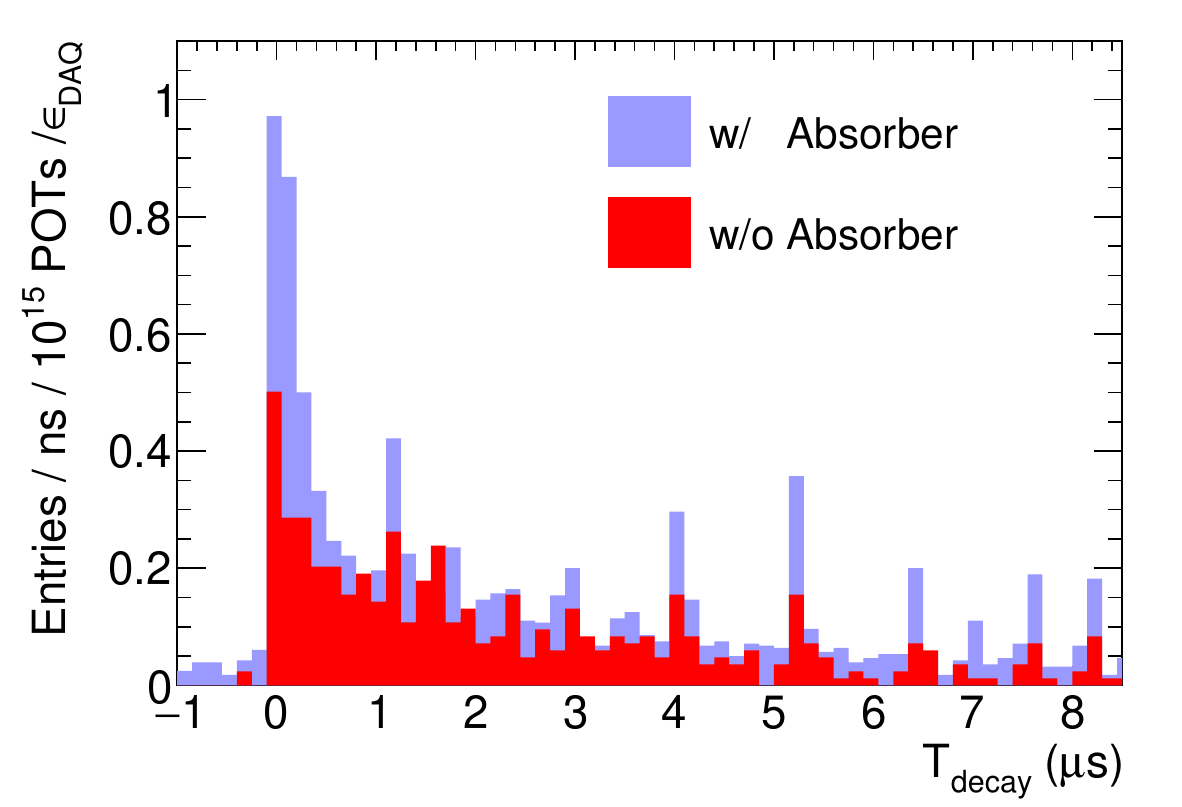}
  \subcaption{Effect of the copper absorber}
\end{minipage}
\caption{
  Decay-time spectra used to assess the effects of the TS magnetic field and copper absorber.
  Each spectrum is normalised by the corresponding number of protons on target and DAQ efficiency.
  (a) The spectra with the TS magnet switched on and off; with the magnet switched off, the measured spectrum is strongly attenuated and no clear exponential component is observed.
  (b) The decay-time spectra with the copper absorber installed and removed; removing the absorber diminishes the short-lifetime component, whereas it is significant when the absorber is installed.
}
\label{fig:Results.ValidationSpectra}
\end{figure}

\fig{Results.ValidationSpectra}(a) presents the decay-time spectra recorded with the TS magnet switched on and off.
With the magnet switched off, the measured spectrum is substantially reduced, and no clear exponential component is observed.
This demonstrates that the signal observed with the magnet on is associated with magnetic transport through the TS.

\fig{Results.ValidationSpectra}(b) shows the corresponding measurements with the copper absorber installed and removed.
Removing the absorber markedly reduces the short-lifetime component to a fitted value consistent with zero, while the long-lifetime component remains at a similar level in the two configurations.
The fitted short-component yields are $\N{short}=0.00^{+12.50}_{-0.00}$ without the absorber and $354.29^{+53.00}_{-50.51}$ with the absorber.
With the absorber installed, a significant short-lifetime component is observed, and the extracted lifetime is consistent with the expected timescale for muonic copper.

Together, these comparisons confirm the signal interpretation used in the main analysis.

\subsection{Degrader scan}
\label{sec:Results.Deg}

\begin{figure}[tb]
\centering
\begin{minipage}[t]{0.48\linewidth}
  \centering
  \includegraphics[width=\linewidth]{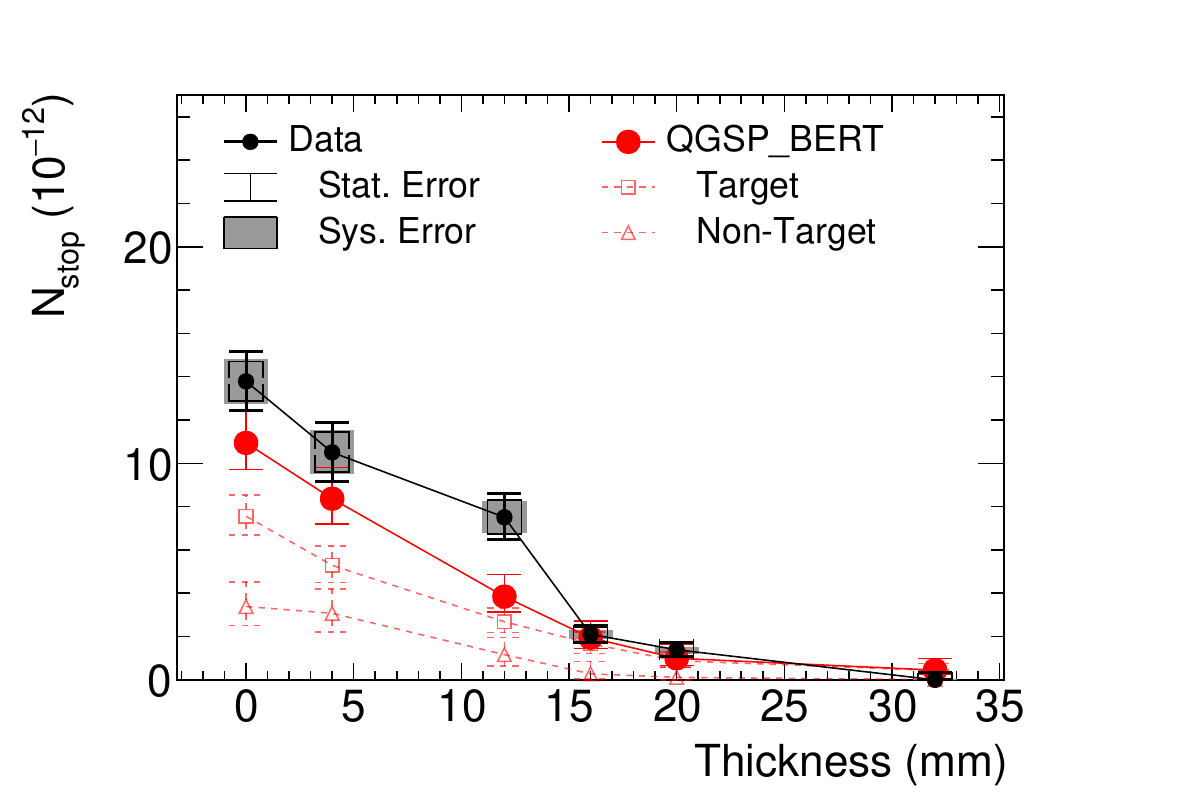}
  \par\smallskip
  {\small (a) Stopped-muon yield $\N{stop}$.}
\end{minipage}
\hfill
\begin{minipage}[t]{0.48\linewidth}
  \centering
  \includegraphics[width=\linewidth]{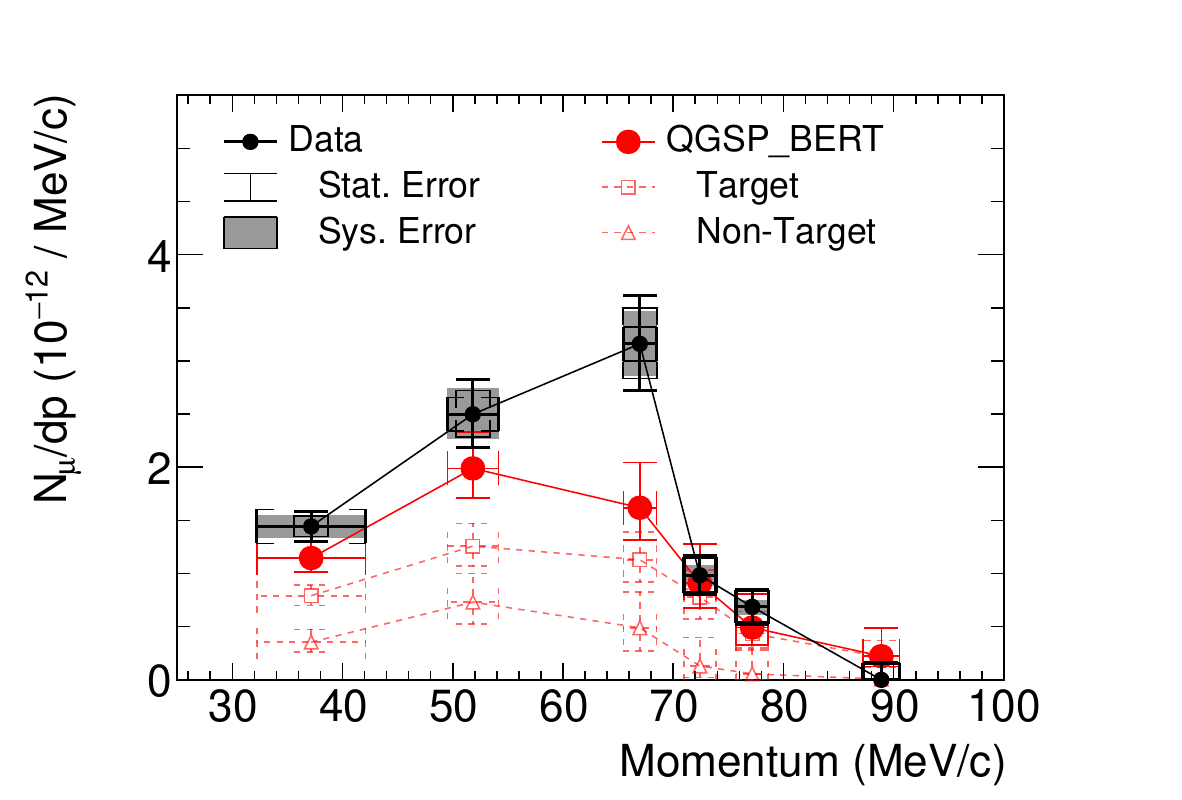}
  \par\smallskip
  {\small (b) Momentum-spectrum estimate $\fdNdp{}$.}
\end{minipage}
\caption{
  Degrader-scan results: (a) the stopped-muon yield $\N{stop}$ and (b) the momentum-spectrum estimate $\fdNdp{}$ obtained from the range-curve conversion of $\N{stop}$.
  For the data, the main error bars show the total uncertainty; the thin error bars and grey boxes show the statistical and systematic uncertainties, respectively.
  The Geant4 QGSP\_BERT simulation shows the target-induced and non-target-induced contributions from the Production Target and downstream materials, respectively, and their sum.
  The simulation uncertainty is dominated by Monte Carlo statistics.
}
\label{fig:Results.DegScanSpectrum}
\end{figure}

The degrader scan probes the momentum spectrum of the transported muon beam.
The reconstructed stopped-muon yield $\N{stop}$ in \eq{Analysis.Nstop} and the corresponding momentum-spectrum estimate $\fdNdp{}$ in \eq{Analysis.Nmu} are listed in \tab{Results.Deg} and shown in \fig{Results.DegScanSpectrum}.
The yield decreases with the total degrader thickness, while the spectrum is concentrated in the low-momentum region, with its largest spectral density between approximately \qty{50}{\MeVc} and \qty{70}{\MeVc}.
Both observables are compared with the corresponding Geant4 simulation.

\begin{table}[tb]
  \centering
  \caption[Summary of the degrader scan results.]{
    Summary of the degrader scan results.
    For each value of the total graphite degrader thickness $d_{\rm deg}$, the reconstructed stopped-muon yield $\N{stop}$ and the corresponding momentum-spectrum estimate $\fdNdp{}$ are listed.
    The first and second uncertainties are statistical and systematic, respectively.
    For $d_{\rm deg}=\qty{32}{\mm}$, the fitted central value is finite but below the displayed precision and is therefore shown as 0.00.
  }
  \label{tab:Results.Deg}
  \newcommand{\statSyst}[4]{\ensuremath{^{+#2}_{-#1}\mathrm{(stat.)}^{+#4}_{-#3}\mathrm{(sys.)}}}
  \normalsize
  \setlength{\tabcolsep}{4pt}
  \renewcommand{\arraystretch}{1.28}
  \begin{tabular}{
	    S[table-format=2.0]
	    S[table-format=2.2]@{\hspace{2pt}}l
	    S[table-format=1.2]@{\hspace{2pt}}l
	  }
    \toprule
    {$d_{\rm deg}$ (\unit{\mm})} & \multicolumn{2}{c}{$\N{stop}\ (10^{-12})$} & \multicolumn{2}{c}{$\dNdp{}\ (10^{-12}/\mathrm{MeV}/c)$} \\
    \midrule
	     0 & 13.78 & \statSyst{0.90}{0.93}{1.01}{1.01} & 1.44 & \statSyst{0.09}{0.10}{0.11}{0.11} \\
	     4 & 10.51 & \statSyst{0.91}{0.93}{0.97}{1.01} & 2.50 & \statSyst{0.22}{0.22}{0.23}{0.24} \\
	    12 &  7.51 & \statSyst{0.77}{0.80}{0.70}{0.73} & 3.16 & \statSyst{0.33}{0.34}{0.29}{0.31} \\
	    16 &  2.10 & \statSyst{0.34}{0.35}{0.20}{0.20} & 0.98 & \statSyst{0.16}{0.16}{0.09}{0.09} \\
	    20 &  1.39 & \statSyst{0.29}{0.30}{0.15}{0.13} & 0.69 & \statSyst{0.14}{0.15}{0.07}{0.06} \\
	    32 &  0.00 & \statSyst{0.00}{0.32}{0.00}{0.00} & 0.00 & \statSyst{0.00}{0.16}{0.00}{0.00} \\
    \bottomrule
  \end{tabular}
\end{table}

The simulation reproduces the overall decrease with degrader thickness and the main spectral-shape feature, although the measured points tend to lie above the simulated yield in the main momentum region of interest.
The \qty{12}{\mm} degrader point, corresponding to a representative momentum of about \qty{67}{\MeVc}, exhibits the most pronounced localised discrepancy within this trend.

For this \qty{12}{\mm} dataset, additional fit-stability checks were performed using an independent fit to the full sample and fits to smaller time-ordered subsets.
All results are consistent with the nominal extraction.

\subsection{Beam-position scan}
\label{sec:Results.Pos}

\begin{figure}[tb]
  \centering
  \includegraphics[width=0.82\linewidth]{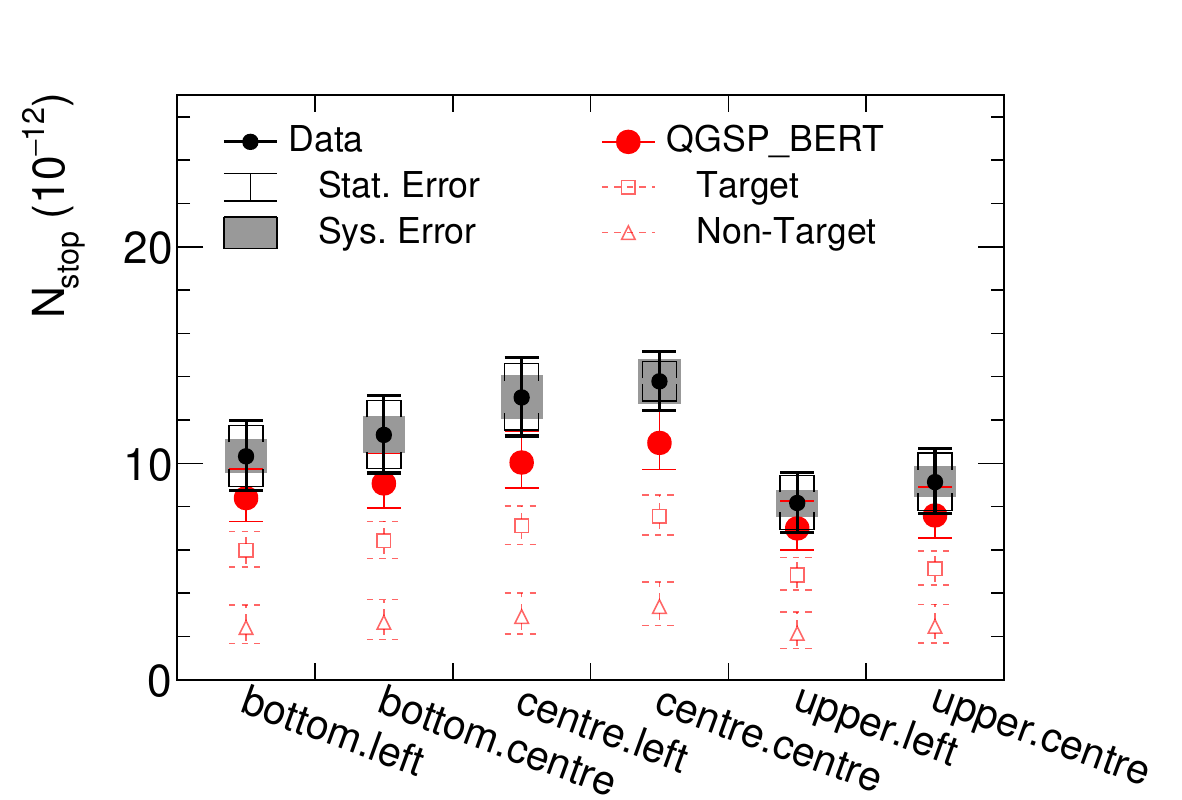}
  \caption{
    Beam-position scan result for the no-degrader configuration, expressed as the stopped-muon yield $\N{stop}$.
    For the data, the main error bars show the total uncertainty; the thin error bars and grey boxes show the statistical and systematic uncertainties, respectively.
    The Geant4 QGSP\_BERT simulation shows the target-induced and non-target-induced contributions from the Production Target and downstream materials, respectively, and their sum.
  }
  \label{fig:Results.PosScan}
\end{figure}

The beam-position scan probes the transverse profile of the transported muon beam.
With no degrader, all six positions probe the same effective momentum band, centred around \qty{37}{\MeVc}.
The reconstructed $\N{stop}$ values from the beam-position scan are listed in \tab{Results.Pos} and compared with the simulation in \fig{Results.PosScan}.

\begin{table}[tb]
  \centering
  \caption{
    Summary of the beam-position scan results for the $d_{\rm deg}=\qty{0}{\mm}$ configuration, where $d_{\rm deg}$ is the total graphite degrader thickness.
    The RC position column denotes the vertical and horizontal detector positions.
    The centre.centre entry is the same dataset as the $d_{\rm deg}=\qty{0}{\mm}$ entry in \tab{Results.Deg}.
    The first and second uncertainties are statistical and systematic, respectively.
  }
  \label{tab:Results.Pos}
  \newcommand{\statSystPos}[4]{\ensuremath{^{+#2}_{-#1}\mathrm{(stat.)}^{+#4}_{-#3}\mathrm{(sys.)}}}
  \normalsize
  \setlength{\tabcolsep}{5pt}
  \renewcommand{\arraystretch}{1.28}
  \begin{tabular}{
    l
    S[table-format=2.2]@{\hspace{2pt}}l
  }
    \toprule
    RC position & \multicolumn{2}{c}{$\N{stop}\ (10^{-12})$} \\
    \midrule
    bottom.left   & 10.32 & \statSystPos{1.38}{1.43}{0.76}{0.79} \\
    bottom.centre & 11.31 & \statSystPos{1.55}{1.60}{0.83}{0.87} \\
    centre.left   & 13.04 & \statSystPos{1.50}{1.56}{0.96}{1.01} \\
    centre.centre & 13.78 & \statSystPos{0.90}{0.93}{1.01}{1.01} \\
    upper.left    &  8.17 & \statSystPos{1.22}{1.27}{0.60}{0.60} \\
    upper.centre  &  9.13 & \statSystPos{1.31}{1.35}{0.67}{0.74} \\
    \bottomrule
  \end{tabular}
\end{table}

The highest yield is observed at the beam centre position.
When the RC is moved away from that position, the yield decreases.
The reduction is asymmetric: moving the detector upward suppresses the yield more strongly than moving it downward, while horizontal displacement to the left has a milder effect.
This pattern is consistent with the expected vertical drift of negative particles in the curved solenoid and its compensation by the dipole field for slow muons.

The simulation captures the central concentration and the vertical asymmetry qualitatively.
To test the sensitivity of this profile to the dipole field that compensates the curvature-induced drift, the simulation was repeated with a field of \qty{40}{\mT}, \qty{10}{\mT} below the nominal setting.
The reduced field shifts the simulated beam centre downward and worsens the agreement with the data.
This comparison supports the interpretation that the measured vertical profile reflects the balance between the curvature-induced drift and its compensation by the dipole field, with the nominal setting providing a better description of the data.

\subsection{Overall comparison between data and simulation}
\label{sec:Results.Comp}

The scan-by-scan comparisons are quantified using a scale factor and a goodness-of-fit statistic.
For the scale factor, the data--simulation difference is expressed in log-ratio space.
For each configuration $i$, $D_i$ and $S_i$ denote the measured and simulated \N{stop} values, and
\begin{equation}
  x_i=\ln\!\left(\frac{D_i}{S_i}\right).
\end{equation}
The scale-factor estimator quantifies any common global offset of the data relative to the simulation across the scan points.
Log-ratio space is chosen to symmetrise the relative scale difference and allow the uncertainties to be treated linearly.
The central value of the log-scale parameter is the weighted mean
\begin{equation}
  \lambda = \frac{\sum_i w_i x_i}{\sum_i w_i}, \qquad
  w_i = \left[
    \frac{\sigma^{+}_{\mathrm{stat}}(x_i)+\sigma^{-}_{\mathrm{stat}}(x_i)}{2}
  \right]^{-2}.
\end{equation}
Denoting the positive and negative statistical uncertainties of $D_i$ and $S_i$ by $\delta D^{\pm}_{i,\mathrm{stat}}$ and $\delta S^{\pm}_{i,\mathrm{stat}}$, the log-ratio uncertainties are
\begin{equation}
  \sigma^{\pm}_{\mathrm{stat}}(x_i)
  \simeq
  \left[
    \left(\frac{\delta D^{\pm}_{i,\mathrm{stat}}}{D_i}\right)^2 +
    \left(\frac{\delta S^{\mp}_{i,\mathrm{stat}}}{S_i}\right)^2
  \right]^{1/2}.
\end{equation}
Here the upper (lower) sign gives the positive (negative) uncertainty on $x_i$.
The quoted statistical uncertainty on $\lambda$ is obtained by repeating the weighted average after shifting each $x_i$ by $\sigma^{+}_{\mathrm{stat}}(x_i)$ or $-\sigma^{-}_{\mathrm{stat}}(x_i)$, with fixed weights.
Systematic uncertainties are propagated in the same way, but are not included in the weights and are quoted separately.
The corresponding multiplicative scale factor is $e^{\lambda}$.

The resulting scale factors are shown in \tab{Results.Comp}.
The nominal rows for the degrader scan, beam-position scan, and combined comparison include all corresponding scan points.
The rows marked with an asterisk repeat the calculation with the $d_{\rm deg}=\qty{12}{\mm}$ dataset excluded.
For the \qty{32}{\mm} degrader point, the finite central value below the displayed precision is used.
All values of $\lambda$ remain positive when the quoted uncertainties are taken into account, indicating that the measured stopped-muon yield is systematically higher than the simulated yield.
The scale factors obtained independently from the degrader scan and the beam-position scan are mutually consistent within uncertainties, and the corresponding values of $e^{\lambda}$ are approximately 1.3, i.e., a scale difference of about \qty{30}{\percent}.

The goodness of fit is also evaluated using $\chi^2$ with the residuals $\Delta_i=D_i-S_i$.
This statistic evaluates the point-to-point consistency once the experimental uncertainties are taken into account.
Two values are quoted: $\chi^2_{\mathrm{stat}}$, which uses only the independent statistical uncertainties, and $\chi^2_{\mathrm{stat+sys}}$, which includes correlated systematic uncertainties through a covariance matrix.
For the latter, the systematic uncertainties in the data and simulation are each treated as fully correlated among datasets within a scan, while correlations between the data and simulation systematic uncertainties are neglected.

The $\chi^2$ values are evaluated using the unscaled simulation central values, and the quoted number of degrees of freedom (ndf) is the number of points compared.
With statistical uncertainties only, the degrader and beam-position scans both give $\chi^2/\mathrm{ndf}$ values larger than unity, with the largest tension in the degrader scan.
Including the correlated systematic uncertainties reduces the corresponding $\chi^2/\mathrm{ndf}$ values to about unity.
Excluding the $d_{\rm deg}=\qty{12}{\mm}$ degrader point further reduces the degrader-scan and combined $\chi^2/\mathrm{ndf}$ values, confirming that this point is the main local contribution to the shape disagreement.

The scale factors and $\chi^2$ values therefore give a consistent picture.
Within the full uncertainties, the simulation captures the main dependence on degrader thickness and detector position, including the qualitative transport behaviour.
The measured stopped-muon yields are systematically higher by about \qty{30}{\percent}.
Together with the scan dependence, this systematic excess provides a quantitative constraint on the absolute low-momentum muon production and transport rate and identifies residual uncertainty in the present production and transport modelling.

Plausible contributors include the production normalisation, TS acceptance, magnetic-field description in the transport region, and completeness of the material and geometry implementation from the Production Target through the beamline.
For non-target-induced components, large structures outside the nominal vacuum transport path, such as the beam dump and surrounding radiation shields, are also relevant.
Small modelling differences in these structures can affect the secondary-particle flux transported to the detector region.

After these potential sources have been addressed, a more detailed study of the physics-list dependence beyond the baseline QGSP\_BERT configuration will be required to assess any remaining discrepancy in the hadron-production model.
The \PhaseAlpha result therefore serves as a validation benchmark and, because the relevant backward secondary-particle production has not previously been directly measured, provides a direct experimental constraint on the production and transport modelling of slow negative muons in COMET.

\begin{table}[tb]
  \centering
  \caption[Summary of the data--simulation comparison.]{
    Summary of the data--simulation comparison for the stopped-muon yield \N{stop}, expressed through the log-scale parameter $\lambda$, the corresponding multiplicative scale factor $e^{\lambda}$, and the goodness-of-fit statistic $\chi^2$.
    The first and second uncertainties on $\lambda$ are statistical and systematic, respectively.
    The number of degrees of freedom (ndf) is the number of scan points compared.
    The $\chi^2/\mathrm{ndf}$ values are shown for statistical uncertainties only and for the combined statistical and systematic uncertainties.
  }
  \label{tab:Results.Comp}
  \normalsize
  \newcommand{\statSystLam}[4]{\ensuremath{^{+#2}_{-#1}\mathrm{(stat.)}^{+#4}_{-#3}\mathrm{(sys.)}}}
  \setlength{\tabcolsep}{2pt}
  \renewcommand{\arraystretch}{1.24}
  \begin{tabular}{
    >{\raggedright\arraybackslash}p{0.22\linewidth}
    S[table-format=1.2]@{\hspace{3pt}}
    l@{\hspace{10pt}}
    S[table-format=1.2]@{\hspace{10pt}}
    S[table-format=2.0]@{\hspace{10pt}}
    S[table-format=2.2]
    S[table-format=1.2]
    S[table-format=2.2]@{\hspace{10pt}}
    S[table-format=1.2]
  }
    \toprule
    \multirow{2}{*}{Compared datasets} & \multicolumn{2}{c}{\multirow{2}{*}{$\lambda$}} & \multicolumn{1}{c}{\multirow{2}{*}{$e^{\lambda}$}} & \multicolumn{1}{c}{\multirow{2}{*}{ndf}} & \multicolumn{2}{c}{Stat. only} & \multicolumn{2}{c}{Stat. + sys.} \\
    \cmidrule(lr){6-7} \cmidrule(lr){8-9}
    & \multicolumn{2}{c}{} & & & {$\chi^2$} & {$\chi^2/\mathrm{ndf}$} & {$\chi^2$} & {$\chi^2/\mathrm{ndf}$} \\
    \midrule
    Degrader scan & 0.32 & \statSystLam{0.11}{0.09}{0.09}{0.10} & 1.38 &  6 & 21.65 & 3.61 & 10.18 & 1.70 \\
    Degrader scan (*) & 0.26 & \statSystLam{0.11}{0.09}{0.09}{0.09} & 1.30 &  5 &  9.25 & 1.85 &  4.77 & 0.95 \\
    Beam-position scan & 0.26 & \statSystLam{0.08}{0.07}{0.07}{0.08} & 1.29 &  6 & 12.95 & 2.16 &  5.36 & 0.89 \\
    Combined & 0.28 & \statSystLam{0.07}{0.06}{0.08}{0.09} & 1.33 & 11 & 30.00 & 2.73 & 10.77 & 0.98 \\
    Combined (*) & 0.25 & \statSystLam{0.07}{0.06}{0.08}{0.08} & 1.29 & 10 & 17.60 & 1.76 &  6.35 & 0.63 \\
    \bottomrule
  \end{tabular}
  \vspace{4pt}

  (*) These evaluations exclude the dataset with a total degrader thickness of \qty{12}{\mm}.
\end{table}
\section{Conclusion}
\label{sec:Conclusion}

The COMET \PhaseAlpha experiment was the first beam operation at the COMET facility, carried out to commission proton beam delivery, operate the TS with beam, and validate slow negative-muon beam transport.
In this paper, we have presented the measurement of that beam with the RC in the experimental area.
Stopped negative muons were identified by detecting delayed DIO electrons emitted after the muons stopped in a copper absorber.
This allowed the stopped-muon yield, the corresponding momentum spectrum, and the transverse beam profile to be reconstructed.

The degrader scan shows that the transported beam is concentrated in the low-momentum region below about \qty{80}{\MeVc}.
The beam-position scan shows a transverse profile qualitatively consistent with the expected behaviour of negative muons in the curved solenoidal field.
These results demonstrate that the TS successfully transported a measurable slow negative-muon beam to the experimental area.
This successful transport is a crucial prerequisite for the Phase-I programme.

The Geant4-based simulation reproduces the main momentum- and position-dependent trends in the data, while the measured yields are systematically higher by roughly \qty{30}{\percent}.
The \PhaseAlpha dataset therefore provides the first direct experimental benchmark of slow negative-muon transport in COMET and a calibration target for improving the beam model, material description, and simulation of particle production and transport.
In particular, it provides a direct experimental constraint on the backward secondary-particle production that supplies the transported beam.
This establishes the practical basis for future beam characterisation studies and for quantitative investigations of secondary particle production in later stages of the COMET programme.
We thank KEK and J-PARC, Japan, for their support of the COMET infrastructure
and the operation of the COMET Phase-$\alpha$ experiment. This work was supported in part by:
the Japan Society for the Promotion of Science (JSPS) KAKENHI Grant Nos. JP22H00139,
JP22K21350, JP22K03657, JP21H04482, JP21H04971, JP20H01936, and JP22H04941;
the Japan Society for the Promotion of Science (JSPS) Core-to-Core Program
under Grant No. JPJSCCA20220003;
the State Program for Scientific Research ``Convergence-2030'' (item 2.1.01);
the National Natural Science Foundation of China under Grant Nos. 12075326, 12105327, 12575207, and E611465Y10;
the Guangdong Basic and Applied Basic Research Foundation under Grant Nos.
2025A1515010669 and 2023B1515120067;
the Natural Science Foundation of Guangzhou under Grant No. 2024A04J6243;
the Beijing Natural Science Foundation, China (International Scientists Project)
under Grant No. IS23012;
the International Partnership Program of the Chinese Academy of Sciences
under Grant No. 113111KYSB20190035;
the Ministry of Education, Science and Youth of Georgia and the Shota Rustaveli
National Science Foundation of Georgia (SRNSFG) under Grant Nos. DI-18-293,
FR-18-11405, and FR-25-095;
the Deutsche Forschungsgemeinschaft under Grant No. ZU 123/24-2;
the National Research Foundation (NRF) of the Republic of Korea under Grant No.
RS-2024-00354342;
the Universiti Teknologi Malaysia (UTM) Matching Grant Q.J130000.3054.04M23;
the Ministry of Science and Higher Education of the Russian Federation;
the STFC Newton--Ungku Omar Fund and the Royal Society;
the European Union's Horizon 2020 Marie Sk{\l}odowska-Curie Actions under Grant Agreement Numbers 101033237 and 101003460;
and the Joint Institute for Nuclear Research (JINR), project COMET \#1144.

Crucial computing support from all partners is gratefully acknowledged, in particular
from the KEK Central Computing System (KEKCC), Japan, and CC-IN2P3, France.

\let\doi\relax
\bibliographystyle{ptephy}
\bibliography{biblio}

\makeatletter
\global\NAT@stdbstfalse
\makeatother

\end{document}